\documentstyle[aps,prl,manuscript]{revtex}
\begin{document}
\title{Phenomenological theory of \\ two-dimensional quantum liquids}
\author{Mikl\'{o}s Gul\'{a}csi}
\address{Department of Theoretical Physics, Institute of Advanced Studies \\
The Australian National University, Canberra, ACT 0200, Australia}
\maketitle
\begin{abstract}
A phenomenological theory is presented for two-dimensional quantum
liquids in terms of the Fermi surface geometry. It is shown that there 
is a one-to-one correspondence between the properties of an interacting 
electron system and its corresponding Fermi surface.  By doing this, the 
concept of Fermi surface is generalized to include different topologies.
It is shown that for a Fermi liquid the corresponding Fermi surface is 
rough. In the presence of a condensate, the Fermi surface is faceted, while 
for a ferromagnetc instability, the Fermi surface becomes a frozen solid. I 
also determine the surface tension, the step free energy, low lying excitations
and other surface and transport properties of the Fermi surface. The different 
transitions between these phases are also determined. A non-Fermi liquid phase 
is shown to be a pre-roughening state of the Fermi surface, the properties of 
which are briefly analyzed.  
\end{abstract}
\vfill
{\sl Appeared in  the Special Issue of Philosophical Magazine B (vol. 74, 
pp 587) - Forty five years of many-body theory - dedicated to David Pines 
in celebration of his 70th birthday.}
\newpage

\section{Introduction}
\label{sec:intro}

I present a completely new semi-classical approach to the quantum
liquid theory in terms of true surface properties and geometry of the
Fermi surface. Within such a description, it can be shown that there is
a one-to-one correspondence between the properties of a two-dimensional
interacting electron system and its one-dimensional Fermi surface.
Contrary to popular belief and expectation, I will also prove that the
Fermi liquid theory corresponds to a rough and not smooth Fermi
surface.  Some of the evidence can be applied to a three-dimensional
interacting electron system.

The bulk of the paper is dedicated to the detailed analysis of the
rough phase and the corresponding roughening transition. For
completeness, I present in Sections\ \ref{sec:surf} and \ref{sec:rough}
the general theory of equilibrium surfaces and the roughening
transition. In addition, the theory of the more particular
Pokrovsky-Talapov transition is presented in the Appendix. This
standard surface physics theory is applied to Fermi surfaces in
Section\ \ref{sec:FS}, while in Section\ \ref{sec:facet} the possible
transition between a rough and a flat faceted Fermi surface is
described. Due to the present interest in non-Fermi liquid theories, I
will also give a possible scenario in Section\ \ref{sec:nonfermi} when
and how such a phase could appear.

\section{Equilibrium Surfaces}
\label{sec:surf}

The theory of the equilibrium surfaces dates back to the celebrated
work of Wulff (Wulff 1901) and by the time of Herring's 1951 paper
(Herring 1951) the theory was truly understood. The equilibrium surface
shape is the shape that minimizes the total free energy of the surface
at a fixed volume. This constrained minimization was originally solved
by Wulff, which is an elegant construction and valued textbook
material. To find the equilibrium surface shape, follow the rules
(Wulff 1901): {\it i}) make a polar plot (the so-called {\it Wulff
plot}) of the surface tension $\alpha (\hat n)$ as a function of $\hat
n$, {\it ii}) starting at the origin, draw the radius in direction
$\hat n$ out to the Wulff plot, then {\it iii}) construct planes
perpendicular to $\hat n$'s through the intersection of the radius with
the Wulff plot. The interior envelope of this family of planes is
always a convex figure whose shape is that of the equilibrium
surface\footnote{The overall magnitude has no significance.}.

In the following, I present the analytic derivation of the equilibrium
surface shape based on the Landau and Andreev derivation (Landau and
Lifshitz 1980), which is not the most general (it relies on $\alpha
(\hat n)$ being piecewise differentiable) but it has the advantage of
being clear and easy to follow.

As noted previously, the Wulff construction always produces a convex
shape, which can be divided into a top and its corresponding
mirror-image bottom surface, with respect to the $+z$ and $-z$
direction. For simplicity, hereafter I only deal with the top surfaces,
as all the obtained results can be mapped to the bottom one. Through
out this paper I will use the notations: $z(x, y)$ is the height of the
surface with slopes $\partial z / \partial x = h (x)$ and $\partial z /
\partial y = h (y)$,  steps on the surface will have height $H$ and
width $L$ with corresponding tilting angle $\theta$. With these
notations the total free energy of the surface is: $\int dx dy f(\hat
n)$, with
\begin{equation}
f(\hat n) ~ = ~ \alpha (\hat n) \: {\sqrt{1 ~ + ~ h^2 (x) ~ + ~ h^2 (y)}} ~~~
\label{surf-egy}
\end{equation}
being the surface free energy per unit area. Minimizing the total
surface free energy keeping the volume fixed is identical (Landau and
Lifshitz 1980) to solving the variational equation
\begin{equation}
\delta \: \int dx dy \: [ f ~ - ~ 2 \lambda z ] ~ = ~ 0 ~~~ ,
\label{surf-ketto}
\end{equation}
where $\lambda$ is a Lagrange multiplier. Eq.\ (\ref{surf-ketto})
transforms into
\begin{equation}
\int dx dy \: \Big[ {\frac{\partial f}{\partial h (x)}} \: {\frac{\partial
\delta z}{\partial x}} ~ + ~ {\frac{\partial f}{\partial h (y)}} \: {\frac{
\partial \delta z}{\partial y}} ~ - ~ 2 \lambda \: \delta z \Big] ~ = ~ 0 ~~~ .
\label{surf-harom}
\end{equation}
By assuming its variation to be equal to zero, for any height variation
$\delta z$, Eq.\ (\ref{surf-harom}) can be partially integrated. The
obtained result, again, must be valid for any $\delta z$, accordingly a
simple integration with a proper choice of the integration constant
gives:
\begin{equation}
{\frac{\partial}{\partial x}} \: {\frac{\partial f}{\partial h (x)}} ~ + ~
{\frac{\partial}{\partial y}} \: {\frac{\partial f}{\partial h (y)}} ~ + ~
2 \lambda ~ = ~ 0 ~~~ .
\label{surf-negy}
\end{equation}
The solution of Eq.\ (\ref{surf-negy}) is (Landau and Lifshitz 1980):
\begin{equation}
f ~ = ~ \lambda \: [z ~ - ~ x h (x) ~ - ~ y h (y)] ~~~ .
\label{surf-ot}
\end{equation}
Eq.\ (\ref{surf-ot}) is non other than the Wulff plane perpendicular to
$\hat n$ and the interface profile will have the form:
\begin{equation}
x ~ = ~ - {\frac{1}{\lambda}} \: {\frac{\partial f}{\partial h (x)}} ~~~ ,  ~~~
y ~ = ~ - {\frac{1}{\lambda}} \: {\frac{\partial f}{\partial h (y)}} ~~~
{\rm and} ~~~
z ~ = ~ {\frac{1}{\lambda}} \: \Big[ f ~ - ~ h (x) \: {\frac{\partial
f}{\partial h (x)}} ~ - ~
h (y) \: {\frac{\partial f}{\partial h (y)}} \Big] ~~~ .
\label{surf-hat}
\end{equation}
The shape of the surface $z (x, y)$ is obtained as a Legendre transform
of the surface tension per unit area, and the surface free energy is
the Legendre transform of the surface shape:
\begin{equation}
f ~ = ~ z ~ - ~ x \: {\frac{\partial z}{\partial x}} ~ - ~ y \:
{\frac{\partial z}{\partial y}} ~~~ .
\label{surf-het}
\end{equation}
Note, that the above construction is so general that it applies to any
equilibrium interface.  As an example I present in Fig.\ 1 the
equilibrium shape of a square with its Wulff plot. It is important to
emphasize that in first approximation one would consider the surface
tension of a square to be vanishing, as the surface radius is
infinitely large.  However, this is not the case. For a proper
evaluation of the surface tension we must count all the broken bonds at
the surface, or simply the number of bonds to which a periodic boundary
condition is applied (number of periodic bonds). Consider, for
simplicity,  that for the square in Fig.\ 1 the interactions in $z$ and
$x$ directions are equal to unity, then the calculated surface tension
is: $\alpha (\hat n) = \vert \sin \theta \vert + \vert \cos \theta
\vert$.

\section{The Roughening Transition}
\label{sec:rough}

The disappearance of a cusp, sharp edges, canonical points, etc from
the Wulff plot are all related to phase transitions. From these, the
most studied is connected to the disappearance of cusps.  Since the
first theoretical attempts by Burton and Cabrera in 1949 (Burton and
Cabrera 1949),  this singularity was usually called the {\it roughening
transition} or the phenomena of {\it surface melting}\footnote{Some
authors use the disordering transition denomination.}. These two
denominations refer to the same thermodynamic transition. Historically
{\it roughening transition} is usually used in the theoretical works
where the emphasis is on the critical phenomena aspect of the phases
transition, while {\it surface melting} is an approach from the point
of view of surface physics and experimentalists. This may generate some
confusion, but everything becomes clear if we follow the argument of
Burton {\sl et al.} (Burton and Cabrera 1949; Burton, Cabrera and Frank
1951) outlined in the following: the static properties of a
two-dimensional surface are well described by a Hamiltonian\footnote{This 
is the well-known Solid-on-Solid model for a crystal-vapor interface, or 
the ''Kossel''-crystal.}
\begin{equation}
H ~ = ~ 2 J \sum_{\langle i, j \rangle} \: \vert H_i ~ - ~ H_j \vert ~~~ ,
\label{BC-egy}
\end{equation}
where the sum runs over all pairs of nearest-neighbour sites. With the
simplest choice for the height variables, $H_i = \{ 0, 1 \}$,
Eq.\ (\ref{BC-egy}) is identical to the two-dimensional Ising model
($S_i = 2 H_i - 1$) which undergoes a second-order phase transition at
$k_B T_c = 2.27 J$, below which $\langle H_i \rangle < 1/2$ and above
which $\langle H_i \rangle = 1/2$. For the surface this means that
above $T_c$ the surface is {\it rough}, as the surface would consist of
50$\%$ each atoms and holes, but the nucleation barrier for
two-dimensional nucleation would vanish, accordingly the surface is
{\it melted}.

The presence of a cusp in the Wulff plot, is due to the appearance of
an absolute value in the expansion of the surface free energy or
surface tension on the smooth side of a surface, e.g., for a
one-dimensional interface:
\begin{equation}
f [h (x)] ~ = ~ f (0) ~ + ~ \vert h (x) \vert \: f_{\rm s} ~ + ~ \ldots ~~~ ,
\label{rough-egy}
\end{equation}
or
\begin{equation}
\alpha (\theta) ~ = ~ \alpha_{0} ~ + ~ \vert \theta \vert \: \alpha_{1} ~ +
\label{rough-ketto}
\end{equation}
respectively. The coefficient $f_{\rm s}$ of $\vert h (x) \vert$ called
the {\it step free energy} is the energy cost of creating a unit length
of step or ledge of unit height.  When $\vert h (x) \vert$ is finite,
adjacent steps are well separated and the energy per unit step length
is well defined. The size of a horizontal facet is proportional to the
jump $\partial f / \partial h (x)$, or equivalently, to the jump in
$\partial \alpha / \partial \theta$ [as $\tan \theta \equiv h (x)$] at
the cusp which follows immediately from Eqs.\ (\ref{surf-harom}) and
(\ref{surf-hat}): when $h (x)$ crosses zero, $x$ jumps from $+ f_{\rm
s} / \lambda$ to $- f_{\rm s}  / \lambda$, indicating that the facet
size is $2 f_{\rm s} / \lambda$.  Accordingly, the roughening
transition is characterized macroscopically by the disappearance of a
cusp from the Wulff plot or governed microscopically by the vanishing
of the step free energy. The Wulff plot will be cusped for
{\underline{facets}} with smooth surface, while the cusp disappears
from the Wulff plot of rough {\underline{rounded}} surfaces.

Usually, the singularity at this transition is in the same universality
class as the Kosterlitz-Thouless (1973) transition. The {\it smooth}
interfacial phase maps to the high-temperature, {\it exponential} $XY$
phase, while the {\it rough} interfacial phase maps to the
low-temperature {\it algebraic} ({\it Gaussian}) $XY$ phase.  Or in
sine-Gordon language, the rough phase maps into a non-interacting, or
Tomonaga-Luttinger model, while the smooth surface into a Luther-Emery
model.  However, there are other types of continuous transitions, which
can give rise to a rough surface, such as Ising, Potts,
Pokrovsky-Talapov (1978)  or chiral that are more frequent, depending
upon symmetry, number of states, interaction parameters, fields,
chirality, etc. An analysis of the critical behaviour of these
transitions is not the purpose of the present work. Such a microscopic
approach to the rough surfaces in general and to the Fermi surface in
particular will be published elsewhere (Gulacsi 1996).

Most real crystals grow under conditions that are far from equilibrium.
The application of the present theory for a real surface is limited by
the requirement of equilibrium. From this point of view the Fermi
surface is an ideal surface as by construction\footnote{The fixed
volume condition is guaranteed by the Luttinger theorem. In most of the
cases through out this work, the Luttinger theorem is local {\sl \'{a}
la} Haldane (Haldane 1994).} it is in equilibrium. 

\section{Surface Properties of Fermi Liquids}
\label{sec:FS}

The Fermi liquid picture is both simple and profound. With a few basic
assumptions it describes accurately the low-energy, long-wave length
behaviour of interacting systems in particular and the thermodynamic
properties of the whole system in general.  The theory assumes that
there is a one-to-one correspondence between the low-energy excitations
of a non-interacting system and the low-energy excitations of the
interacting system while preserving the quantum numbers\footnote{For
simplicity in the following and through out the present work only the
spinless case will be discussed.}.  That is, if the non-interacting
system is characterized by some distribution function $n({\bf k})$ of
the bare particles, by switching on the interaction adiabatically the
eigenstates of the interacting system, called quasiparticles, are also
described by same function $n({\bf k})$. Accordingly, the excitation of
the system is simply measured by the departure $\delta n({\bf k})$ from
the ground-state distribution:
\begin{equation}
\delta n({\bf k}) ~ = ~ n({\bf k}) ~ - ~ n^{0}({\bf k}) ~~~ .
\label{egy}
\end{equation}

For the ideal system,  there exists a simple linear relation between
the energy of a given state and the corresponding distribution
function. When particle interaction is taken into account, the relation
between the state energy, $E$, and the quasiparticle distribution
function, $n({\bf k})$, becomes much more complicated. It may be
expressed in a functional form, $E [ n({\bf k}) ]$, which one cannot in
general specify explicitly.  If, however, $n({\bf k})$ is sufficiently
close to the ground-state distribution $n^{0}({\bf k})$, we can carry
out a Taylor expansion of this functional. On writing $n({\bf k})$ in
the form Eq.\ (\ref{egy}) and taking $\delta n({\bf k})$ to be small,
or to extend over a small region in momentum space, the Landau free
energy (Landau 1957) becomes:
\begin{equation}
F ~ - ~ F_{0} ~ = ~ \sum_{\bf k} [\epsilon ({\bf k}) - \mu] \: \delta
n({\bf k}) ~ + ~
\frac{1}{2} \sum_{{\bf k}, {\bf k}^{\prime}} f({\bf k}, {\bf k}^{\prime}) \:
\delta n({\bf k}) \: \delta n({\bf k}^{\prime}) ~ + ~ \ldots ~~~ ,
\label{ketto}
\end{equation}
which is the heart of the phenomenological theory  of Fermi liquids
(Pines and Nozi\`{e}res 1988) with $f({\bf k}, {\bf k}^{\prime})$
representing the effective interaction between quasiparticles.

It is not my purpose to further analyze Eq.\ (\ref{ketto}) from the
Fermi liquid point of view\footnote{The excitation of
Eq.\ (\ref{ketto}), i.e. the sound modes and the particle-hole
continuum, will be briefly analyzed in Section\ \ref{sec:concl}}, but
only to establish its surface properties as a semi-classical interface.
For simplicity, I restrict ourselves to one-dimensional Fermi
surfaces\footnote{The surface under study, being defined in the
momentum space, the {\it local} curvilinear surface coordinate is
$k_x$, $k_z (k_x)$ denotes the {\it local}  height of the surface with
slope $\partial k_z / \partial k_x = h (k_x)$. The tilting angle
remains $\theta$. It is also important to emphasize that the steps are
associated with the world lines of fermions (Pokrovsky-Talapov 1978).}
(two-dimensional systems). It should be noted that some formulas or
demonstrations that appear in the text are also valid or easily
generalized for a two-dimensional Fermi surface (three-dimensional
systems). The reader will be prompted when these examples arise.

Based on the definitions presented in Sections\ \ref{sec:surf} and
\ref{sec:rough} it is immediate that Eq.\ (\ref{ketto}) describes a
{\it rough} surface because the step free energy is zero\footnote{By
construction Eq.\ (\ref{ketto}) describes density fluctuations of a
liquid surface, rather than elastic waves propagating on a smooth
facet.}.  As noted already by Pines and Nozi\`{e}res (1988) Eq.\
(\ref{ketto}) looks like a Taylor expansion, but it is not. With the
notations from the previous sections and only if $\mu$ and $\delta
n({\bf k})$ are finite, $F - F_{0}$ defined in Eq.\ (\ref{ketto}) is of
the order $h^2 (k_x)$\footnote{See, Eq.\ (\ref{facet-het}) for an
expansion of a rough surface.} because $\epsilon ({\bf k}) - \mu
\propto h (k_x)$ and $\delta n({\bf k}) \propto h (k_x)$.  Accordingly,
the corresponding step free energy, $f_{\rm s}$, is strictly zero and
by definition the surface is rough. The roughening transition is
defined by the vanishing of the step free energy, and any smooth
surface must characterized by Eq.\ (\ref{rough-egy}). The only case
where such a simple proof would not work is the case $\mu = 0$, which
would correspond to a half-filled band systems, a case which I will
analyze later.

This proof should work equally well for two-dimensional Fermi surfaces
for the non-interacting case. However, some problems will appear if we
consider the two-dimensional Fermi surface, a quantum interface case,
which is analyzed in detail by Gulacsi (1996)\footnote{Fisher and Weeks
(1983) and Fradkin (1983) and others have contested the Andreev and
Pashkin (1978) argument about the roughness of the quantum liquid
surfaces at zero temperature.  However, new arguments appear in the
favor of the roughness, namely as presented in Footnote\ 14, the width
of the steps on a quantum surface at $T \rightarrow 0$ are very large,
so the quantum effects are becoming smaller, accordingly the quantum
surface becomes classical before the complete delocalization of the
interface.}.  For a one-dimensional Fermi surface, the above argument
is based on a nearest-neighbour type effective interaction.  However,
at large length scales the discreteness of the step heights is
irrelevant, the surface always can be described by a Gaussian model,
having the properties of a transverse vibrating elastic network. Due to
this, the obtained transition (for details see,
Section\ \ref{sec:facet}) is {\it not destroyed} if the interaction
between quasiparticles is of short-range type or even has the form
$f({\bf k}, {\bf k}^{\prime}) \sim g/\vert k^{}_{x} - k^{\prime}_{x}
\vert^{\gamma}$ with $\gamma \le 2$ and $g > 0$. From these
interactions the $g / \vert k^{}_{x} -  k^{\prime}_{x} \vert^2$ form is
the most probable one (Ashcroft and Mermin 1976) based on a
Hartree-Fock theory, or simply because the elastic or bipolar
interactions between steps on an equilibrium surface have this form
(see, Section\ \ref{sec:facet}).

The Fermi surface is indeed rough for the non-interacting case, and it
can be directly proven by the Wulff plot.  Equations\ (\ref{surf-egy})
through Eq.\ (\ref{surf-het}) enable one to reconstruct the Wulff plot
from the equilibrium surface shape. For an isotropic and
non-interacting system with $\epsilon_F = \hbar^2 \vert {\bf k}_F
\vert^2 / 2m$ we immediately obtain that $\alpha (\theta) = \epsilon_F$
becomes independent of $\theta$, as in the case for an {\it ordinary
fluid}. Accordingly it has no cusps; that is, the Fermi surface is
rough. In this simple case, following Nozi\`{e}res (1991), we can
determine the Lagrange multiplier of the Fermi surface appearing in
Eqs.\ (\ref{surf-hat}): $\lambda  = v_F / 4 m$.

On a lattice\footnote{For simplicity consider a square lattice.} the
half-filled and doped case must be treated separately. Namely, for
$H_{0} = -t \sum_{\langle i,j \rangle} (c^{\dagger}_{i} c^{}_{j} + {\sl
h.c.}) + \mu \sum_{i} c^{\dagger}_{i} c^{}_{i}$, with $\mu = -2 t
\delta$, where $\delta$ denotes the hole doping, using
Eq.\ (\ref{surf-het}), the Wulff plot corresponding to the Fermi
surface $\mu = 2t \sum_{i} \cos k_{i}$ will be smooth for $\mu \ne 0$
and cusped for $\mu = 0$.  A similar argument used in the previous
paragraphs is that $\delta n({\bf k}) \propto h (k_x)$ and also
$\epsilon ({\bf k}) - \mu \propto h (k_x)$ and for $\mu \ne 0$, i.e.,
$F - F_{0} \propto h^2 (k_x)$, while for $\mu = 0$ only $\delta n({\bf
k}) \propto h (k_x)$, so $F - F_{0} \propto h (k_x)$, accordingly the
step free energy is finite. In this case, the Fermi surface and its
Wulff plot are identical to Fig.\ 1 (rotated by $\pi / 2$).  That is,
the Fermi surface is faceted and the facets are smooth. However, in
this case, the appearance of the facets can be attributed to the
presence of the van Hove singularity.  The step free energy is finite
but small close to the van Hove singularity, increasing as we approach
the singularity.  This behaviour is caused by the vanishing group
velocity, $\vert \nabla \epsilon ({\bf k}) \vert$, which explicitly
shows up in evaluating Eq.\ (\ref{surf-het}). The step free energy,
however, cannot grow infinitely large, is bounded by the crystal
structure, but produces a facet with the largest possible flat surface,
and this is the Fermi surface of the half-filled band case.  The above
arguments for the half-filled band case are also valid for a
two-dimensional Fermi surface.

In conclusion, in a non-interacting system the change of the topology
of the Fermi surface, i.e. the appearance of flat facets can be
attributed to the presence of the van Hove singularity.  For an
interacting case it is difficult to determine the Wulff plot because we
do not know the exact form of the self energy, $\Sigma_{\rm Re} ({\bf
k}, \omega)$.  However, it can be shown that the Fermi liquid picture
in general, and Eq.\ (\ref{ketto}) in particular, is not valid close to
half filling. For example, for the two-dimensional Hubbard model,
second-order perturbation theory gives for the $f({\bf k}, 
{\bf k}^{\prime})$ function (Honner 1995):
\begin{equation}
f({\bf k}, {\bf k}^{\prime}) ~ = ~ {\frac{4 U^2 \Delta k}{N t \pi^2
{\sqrt{16 \sin^2 \Delta k - \Delta^2 \epsilon}}}}
\: \ln \Big| \: {\frac{ 4 \sin \Delta k + {\sqrt{16 \sin^2 \Delta k -
\Delta^2 \epsilon}}}
{4 \sin \Delta k - {\sqrt{16 \sin^2 \Delta k - \Delta^2 \epsilon}}}} \:
\Big| ~~~ ,
\label{harom}
\end{equation}
where $2 \Delta k = k^{}_{x} - k^{\prime}_{x} = k^{}_{y} -
k^{\prime}_{y}$ and $\Delta \epsilon = \cos k^{\prime}_{x} - \cos
k^{}_{x} + \cos k^{\prime}_{y} - \cos k^{}_{y}$. The $f({\bf k}, {\bf
k}^{\prime})$ from Eq.\ (\ref{harom}) is bounded only in the vicinity
of $(\pi/2, \pi/2)$ with a maximum value of $2 U^2 / N t \pi^2$ and is
divergent as we approach the van Hove singularity. That is, the
adiabatic switching on the interaction picture of the Fermi liquid
theory is no longer valid. 

For an interacting electron system, the half-filled band case is mostly
due to perfect nesting, in the presence of which the Fermi surface
becomes unstable to the formation of long-range order (e.g.,
charge-density-wave or spin-density-wave), or Mott insulator, etc., due
to which a charge or spin gap (depending on the interactions) will
open. This causes the divergence of the $f({\bf k}, {\bf k}^{\prime})$
function. In this case, the Fermi surface is not defined in the sense
of Eq.\ (\ref{egy}) or (\ref{ketto}). Within the present approach,
based on the Fermi surface geometry, we can generalize its definition
with the understanding that only its topology changes, which means that
the Fermi surface becomes faceted.  Previously, the Fermi surface was
rounded, while now it has flat, faceted regions.  The presence of a
facet would immediately indicate that the Fermi surface cannot be
excited by any small amount of energy. A minimum energy is required
(the step free energy defined in Section\ \ref{sec:surf}) to create
steps on a flat and smooth facet and this minimum energy signals the
presence  of a gap in the excitation spectrum.  In the next Section the
width of the facet is determined, in first approximation, to be
proportional to $\Delta / v_F$, where $\Delta$ is a mean-field gap and
$v_F$ is the non-interacting Fermi velocity. Accordingly, the
appearance of these facets on the Fermi surface is equivalent to an
instability of the Fermi surface by which a condensate will emerge in
the interacting electron system. The presence of facets in a
two-dimensional Fermi surface will equally mean an instability of the
Fermi surface towards a quasiparticle gap opening.

\section{The Faceting Transition}
\label{sec:facet}

The half-filled band analyzed in the previous Section is a particular
case where all the Fermi surface is flat. Generally, facets appear on a
surface gradually, meaning their width increases with variation of some
parameters (doping, external magnetic or electric field) or
temperature. In any case, the appearance of facets on the Fermi surface
will always lead to nesting (due to the symmetries of the Brillouin
zone) usually a partial one. As the surface of the facets increases,
nesting becomes stronger and stronger until it reaches a completely
nested Fermi surface similar to the half-filled band case.  The
transition from the rough surface into a faceted one, is a roughening
transition by which the topology of the Fermi surface changes
completely, and the properties of which are determined in the
following.

The surface free energy of a facet is given in Eq.\ (\ref{rough-egy}).
However, this term represents only non-interacting steps. To describe
properly the faceting transition we need the knowledge of the
interactions between the steps present on the surface. As the present
discussion is a phenomenological one, we do not intend to subtract
these interactions from perturbation theories but rather to analyze
them from a surface physics point of view.  There are several main
sources of interaction on a surface. The first is the fact that we do
not allow steps to cross, or create overhangs, as these states would
require higher energy.  Meaning, the steps will be restricted resulting
in an entropic interaction.  The contribution of such an entropic
repulsion, we can understand by following Gruber and
Mullins\footnote{The original proof was for a terrace-step-kink (TSK) 
type step.} (1967):
let us suppose for a moment that a single step is constrained to move
between two walls situated at $k_x = \pm k_L$. The presence of the
walls produces an entropic repulsion between the wall and step. The
probability density $P(n)$ for the step passing true $k_x = n$ is
(Gruber and Mullins 1967):
\begin{equation}
P(n) ~ = ~ {\frac{1}{k^2_L}} \: \cos^2 {\frac{\pi n}{2 k_L}} ~~~ ,
\label{facet-egy}
\end{equation}
and the free energy of the step becomes\footnote{If $P(n)$ would be a
constant, than the step staying away from the walls would decrease its 
entropy and thereby its free energy with $1/L$.}:
 $f_{\rm s} = f_{\rm s} (0) + {\rm const} \: / k^2_L$. Now, replacing
 the fixed boundaries of the walls by neighbouring steps results in:
$f_{\rm s} = f_{\rm s} (0) + {\rm const} \: h^2 (k_x)$, which
introduced in Eq.\ (\ref{rough-egy}) will give an $\vert h (k_x) \vert
\: h^2 (k_x)$ contribution.

Short-range interactions (Jayaprakash, Rottman and Saam 1984) cannot
alter this behaviour.  Namely, in a mean-field theory the contribution
of a short-range interaction $V(k_x)$ will be $(1/L) \int dk_x V(k_x) h
(k_x) G(k_x)$, where $G(k_x)$ is the non-interacting pair correlation
function, $G(k_x) = h (k_x)  \{ 1 - \sin^2 [\pi k_x h (k_x) / L] / [\pi
k_x h (k_x) / L]^2 \} $. If $V$ is smaller than the distance between
steps or the step width\footnote{The step width in the $T \rightarrow
0$ limit can be related (Balibar, Guthmann and Rolley 1993) to the
so-called {\it pinning} or {\it coupling strength} of an effective
sine-Gordon model, accordingly the steps are relatively wide.},
$G(k_x)$ may be expanded and the contribution of a short-range
interaction will be of order $h^4 (k_x)$. 

Other interactions (see e.g., Beijeren, van and Nolden 1987), e.g.,
elastic or dipole interactions, assume the $g / \vert k^{}_{x} -
k^{\prime}_{x} \vert^2$ form on a surface with free energy contribution
in the continuum limit with $P(k_x)$ given in Eq.\ (\ref{facet-egy}),
true the whole length of a facet at small $h (k_x)$ is: 
\begin{equation}
\sim \: \int^{+ (2 \vert h (k_x) \vert)^{-1}}_{- (2 \vert h (k_x)
\vert)^{-1}} dk_x \:
{\frac{\vert h (k_x) \vert}{k_x  +  1 / 2 \vert h (k_x) \vert   }} \:
\cos^2 \pi {k_x} h (k_x) ~ \sim ~ \vert h (k_x) \vert \: h^2 (k_x) ~~~ .
\label{facet-ketto}
\end{equation}
Accordingly, on a facet the surface free energy from Eq.\ (\ref{facet-egy})
will have the form:
\begin{equation}
f [h (k_x)] ~ = ~ f (0) ~ + ~ f_{\rm s} \: \vert h (k_x) \vert   ~ + ~
f_{\rm int.} \: \vert h (k_x) \vert \: h^2 (k_x) ~ + ~ \ldots ~~~ ,
\label{facet-harom}
\end{equation}
where in $f_{\rm int.}$ we can accumulate the effects of all the interactions
present on the surface.

At this point, it is important to emphasize the following: {\it i}) All
type of interactions on a surface are giving a contribution of $h^3_x$
in the expression of the surface free energy.  Naively, one might
expect the free energy due to interactions to be of the $h^2 (k_x)$
order, however it is not. A contribution of the $h^2 (k_x)$ type occurs
only in a rough surface, see Eq.\ (\ref{facet-het}) or
Eq.\ (\ref{ketto}) and the discussions  following it. {\it ii}) The
presence of the interaction term in Eq.\ (\ref{facet-harom}) is crucial
for the faceting transition. The $f_{\rm int.} = 0$ corresponds to a
mean-field type critical bahaviour (see e.g., Rottman and Wortis 1984) 
which may containe a 
tricritical point\footnote{A tricritical point appears for an
attractive interaction and bound states for a repulsive one.}.

In the following I determine the interface profile. Continuing the
arguments from Section\ \ref{sec:rough}, when $h (k_x)$ crosses zero,
the horizontal width of the facet, $k_x$, jumps from $+ f_{\rm s} /
\lambda$ to $- f_{\rm s} / \lambda$, that is the facet width will be
proportional to $f_{\rm s} / \lambda$. In a mean-field approach the
spectrum has the ${\sqrt{\epsilon^2 ({\bf k}) + \Delta^2 ({\bf k})}}$
form. Using the continuum result from Section\ \ref{sec:FS} for the
Lagrange multiplier: $\lambda = v_F / 4$, in first approximation it is
obtained that the Fermi surface facet width is proportional to $\Delta
/ v_F$. In Section\ \ref{sec:nonfermi} it will be shown that the
facet width is inverse proprotional to the 
effective correlation length on the surface: $\xi^{-1} \propto 
f_{\rm s} / \lambda \approx \Delta / v_F$.

The interface profile near the edges of a facet will have a universal
shape. Using
Eqs.\ (\ref{surf-hat}) and (\ref{facet-harom}) one obtains:
\begin{eqnarray}
k_z ~ & = & ~ k^0_z \: \Big[ 1 ~ - ~ {\frac{2 f_{\rm int.}}{f(0)}} \: \vert
h (k_x) \vert^3 \Big]
~~ , \label{facet-negy} \cr
k_x ~ & = & ~ \pm \: k^0_x \: \Big[ 1 ~ + ~ {\frac{3 f_{\rm int.}}{f_{\rm
s}}} \: h^2 (k_x) \Big]
~~ , \label{facet-ot}
\end{eqnarray}
where the $k^0_z = f(0) / \lambda$ and $k^0_x = f_{\rm s} / \lambda$
notations were used.
That is,
\begin{equation}
k_z ~ - ~ k^0_z ~ = ~ - {\frac{2}{3 {\sqrt{f_{\rm int.}}}}} \:
\vert k_x ~ - ~ k^0_x \vert^{3/2} ~~~ .
\label{facet-hat}
\end{equation}
This equation establishes a universal (non-analytic) jump of the
curvature, well-known in surface physics. The $3/2$ exponent is the
Pokrovsky-Talapov (1978) exponent while the transition is called the
Gruber-Mullins (1967) or the Pokrovsky-Talapov transition. The above
description of the Pokrovsky-Talapov transition is mainly based on an
effective nearest-neighbour model, however, as previously demonstrated,
the Pokrovsky-Talapov exponent and correspondingly the transition is
not destroyed by short-range interactions and not even by long-range
interactions. The effects of the long-range interactions are analyzed
in more detail the following Section, where it will be seen that the
roughening transition remains unchanged, only a new phase may appear.

On the rough side of the surface, however, a simple Taylor expansion
cannot be performed. As presented in the Appendix; on the rough side
the free energy will always have the form:
\begin{equation}
f [h (k_x)] ~ - ~ f (0) ~ \propto ~  h^2 (k_x)  ~ + ~ \ldots ~~~ .
\label{facet-het}
\end{equation}
similar to Eq.\ (\ref{ketto}).

Other characteristics of the roughening transition can be re-written
for the Fermi surface, e.g., the universal jump in the curvature of the
Fermi surface before and after the transition, or it can be shown that
near the edge of the facets on the rough curved surfaces the steps are
spontaneously generated with a square root density.

As presented in the Appendix, a Pokrovsky-Talapov transition appears by
varying the a doping, or external magnetic or electric fields. As a
function of temperature, however, such a transition is a
Kosterlitz-Thouless transition if the width of the facet vanishes at
the transition, otherwise is a standard second-order phase transition.

\section{Fermi and Non-Fermi Liquids}
\label{sec:nonfermi}

The conclusions from the previous Sections were that the rough phase of
the Fermi surface corresponds to the metallic (Fermi liquid) state of a
system, while a faceting (Pokrovsky-Talapov type) transition will lead
to an instability of the Fermi surface corresponding to the appearance
of some kind of condensate. In both cases, the surface may be
characterized by the density (occupation probability) distribution
$\rho ({\bf k}) \propto \delta n({\bf k})$, where $\delta n({\bf k}) $
was  defined in Eq.\ (\ref{egy}). As mentioned in
Section\ \ref{sec:surf}, in the rough phase the density-density
correlation function has an algebraic decay but  an exponential decay
for the smooth faceted phase.  It is important to emphasize that these
correlation functions appear in both translational and orientational
order of the surface, which are decoupled (Haldane 1994), with the
understanding that the orientational fluctuations cannot change the
shape of the surface.

The disordered (rough) phase is a non-interacting or Tomonaga-Luttinger
liquid, that is
\begin{equation}
\langle \rho (k_x) \rho (0) \rangle ~ = ~ {\frac{\rm K}{\pi^2 \: k^2_x}} ~ - ~
{\rm K}^{\prime} \: {\frac{\cos [ \pi \: k_x \: h (k_x) ]}{k^{\gamma}_x}} ~~~ ,
\label{nonfermi-egy}
\end{equation}
where ${\rm K}$ is the well-known roughness constant. Equation\
(\ref{nonfermi-egy}) can be viewed as the {\it local slope operator}
so, the height-height correlation function
\begin{equation}
\langle [ h(k_x) - h(0)]^2 \rangle ~ = ~ 2 \int^{k_x}_{0} dk^{\prime}_x \:
(k_x - k_y) \: \langle \rho (k_x) \rho (0) \rangle ~~~ ,
\label{nonfermi-ketto}
\end{equation}
will have the well-known form:
\begin{equation}
\langle [ h(k_x) - h(0)]^2 \rangle ~ \approx ~ {\frac{ 2 \: {\rm K}}{\pi^2}}
\: \ln k_x ~ + ~ {\frac{{\rm K}^{\prime}}{2 h^2_x}} \:
{\frac{\cos [ \pi \: k_x \: h (k_x) ] }{k^{\gamma}_x}} ~~~ ,
\label{nonfermi-harom}
\end{equation}
with the universal constant ${\rm K}$ taking the ${\rm K} = 1$ value,
corresponding to the notations used in Eqs.\ (\ref{nonfermi-egy}) and
(\ref{nonfermi-ketto}). It can be seen from Eq.\ (\ref{nonfermi-egy})
that the non-interacting part adequately describes the properties of
surface. The second term of Eq.\ (\ref{nonfermi-egy}) which is the
characteristic contribution of the Tomonaga-Luttinger models, is not
affecting the height-height correlation function. Accordingly, the
low-energy and long-wave length properties of the metallic phase are
non-interacting in nature. By this we reconfirmed the Fermi liquid
picture.

The faceted phase on the other hand, maps into a Luther-Emery phase:
\begin{equation}
\langle \rho (k_x) \rho (0) \rangle ~ \propto ~ \exp \Big( -
{\frac{k_x}{\xi}} \Big) ~~~ ,
\label{nonfermi-negy}
\end{equation}
where $\xi \propto \lambda / f_{\rm s} \approx v_F / \Delta$ is the
effective correlation length on the surface and $\Delta$ the
quasiparticle gap. The $f_{\rm s}$ is step free energy  defined in
Section\ \ref{sec:FS}. In Section\ \ref{sec:facet}, the above result
was previously derived from a different point of view, namely using
surface minimization arguments and determining the Lagrange multiplier
of the surface.

In the orientational order, i.e. in the $k_z$ direction, the situation
is not so simple and detailed microscopic calculations are needed
(Gulacsi 1996). However, we know that in the metallic phase (rough
surface) there must be a step singularity\footnote{The presence of
singularities as we cross a surface is not a surprise. In the case of a
standard solid-liquid (or solid-gas) interface the mass density has a
singularity at the surface $\delta \rho = \rho_{\rm solid} - \rho_{\rm
liquid}$, which determines the Lagrange multiplier defined in
Section\ \ref{sec:surf} (Nozi\`{e}res 1991).} in $n ({\bf k}) \propto \rho
({\bf k})$ of the $n(k) = \theta (k - k_F)$ form with $\theta (x)$
being the step function. While, in the smooth phase of the Fermi
surface there are no singularities, e.g. in a mean-field description
$n (k) \approx 1 / [\Delta^2 + v^2_F k^2]$.

Concerning a non-Fermi liquid type behaviour, it seems (based on the
above presented analysis) that there are no other phases left.  The
exact solutions of nearest-neighbour interacting lattice models, or
surface models, exhibits three phases (see, Appendix).  First, one
which  is ferroelectric, corresponding to a completely {\it locked} (or
frozen) solid surface,  corresponding to a ferromagnetic type
transition, i.e., a Pomeranchuk (1958) type instability of the Fermi
surface\footnote{The frozen ferroelectric phases of the vertex models
are the only ones that exhibit true long-range order corresponding to a
ferromagnetic state.}. The second, disordered phase and the third,
ordered phase corresponds to the rough and flat (faceted) surfaces,
respectively.  In this context, one of the most appealing possibilities
to generate a liquid state, that is a metallic phase, but not a Fermi
liquid, would be a phase {\underline{intermediate}} between the rough
and flat surfaces, {\it which shares the properties of both}. That is,
a new phase between these two such that the new surface would be
characterized by an exponential decay of the translational order but an
algebraic decay of the correlation function of the orientational order.
The existence of such a surface (sometimes called a hexatic fluid) is
still controversial in the surface community (Bauer 1987), however, it
would correspond to the disordered flat surface introduced by den Nijs
(1989) which is in a pre-roughening state, corresponding to the
presence of a "Haldane gap" type gap. Being so, it has what we could
call {\it hidden} topological order.

The appearance of such a phase is due to short-range interactions
(Nijs, den 1989). In Section\ \ref{sec:facet}, I have shown that
next-nearest-neighbour or short-range interactions cannot disturb or
change the Pokrovsky-Talapov melting transition.  However, what they
can do, is to generate the above introduced disordered flat surface
(Nijs, den 1989).

An algebraic decay of the orientational order, means that $n ({\bf k})$
will have some singularity as we cross $k_F$, and the surface would
correspond to a metallic phase.  However, the metallic phase is not a
Fermi liquid because the translational order is gapped. The step free
energy is zero in such a non-Fermi liquid phase (Nijs, den 1989), so
based on the definitions introduced in Section\ \ref{sec:rough} the
surface is "rough", but the height-height correlations are not
divergent and thus the reason why such a phase is refered to as
pre-roughening phase.

Concerning the translational order, which is gapped, the surface
exhibits flat surface facets similar to the previously analyzed faceted
phase. The main difference is that, in the later phase the facets have
a well-defined macroscopical width (see, Section\ \ref{sec:rough}),
while in a pre-rough phase only the {\it average} surface is flat.  The
surface structure is characterized by perfect antiferromagnetic order
of the heights $h (k_x)$ at random positions $k_x$. Clearly, the
surface structure is complex; that is why it is so difficult to
describe analytically a non-Fermi liquid in higher dimensions. Because
of the random positions of $k_x$, no nesting can occur on the Fermi
surface, accordingly no long-range order, or condensate either, and
indeed, the system will be metallic. The step free energy being zero,
Eq.\ (\ref{ketto}) is still valid, with a specific $f ({\bf k}^{}, {\bf
k}^{\prime})$.  Within the present phenomenological approach the exact
form of $f ({\bf k}^{}, {\bf k}^{\prime})$ cannot be determined.
However, some insights can be gained following a simple analysis: for a
Fermi liquid, as mentioned previously, both orientational and
transversal correlations are algebraic. Being so, there is a
singularity in $n (k)$ and $Z = \{ 1 - \partial \Sigma_{\rm Re} ({\bf
k}, \omega) [\omega = \epsilon ({\bf k})] / \partial \omega \}^{-1}$ is
finite.  Let us consider the pre-rough, non-Fermi liquid, phase.  As
previously discussed, the orientational order is still algebraic; some
sort of singularity will appear in $n (k)$ as we cross $k_F$. The
translational order being gapped $Z$ is zero. So, the emerging
non-Fermi liquid will have a singularity similar to a
Tomonaga-Luttinger liquid in one-dimension. However, I must emphasize
that the above analysis is approximate within a semi-classical approach
to the Fermi surface, and a microscopic treatment is needed (Gulacsi
1996)  to prove exactly the above statement.

\section{Conclusions}
\label{sec:concl}

Based mainly on a semi-classical approach, I have demonstrated that
there is a one-to-one correspondence between the properties of a
two-dimensional interacting electron system and its one-dimensional
Fermi surface. Though, some of the evidence can be generalized to a
three-dimensional interacting electron system.

In a generic scenario, the Fermi surface can have three distinctive
phases, namely the locked solid, which would correspond to a
ferromagnetic type instability, a rough phase, which I have proven to
be the metallic or Fermi liquid phase and a flat faceted phase, which
is an instability of the Fermi surface against the appearance of a
quasiparticle gap. By varying the interaction strength, external
magnetic or electric, fields, or doping, we can cross these phases in
the previously presented order, that is, frozen solid
$\Longleftrightarrow$ rough, melted surface $\Longleftrightarrow$ flat,
faceted surface. It is interesting to note, that there is no transition
between frozen solid and flat faceted surface. The transitions between
these phases are: {\it i}) first or second-order from the frozen solid
to rough phase.  {\it ii}) Kosterlitz-Thouless or Pokrovsky-Talapov
melting between the flat faceted and rough surfaces. The properties of
the rough and faceted surfaces and the transition between these two
phase were described in great detail in Sections\ \ref{sec:surf} -
\ref{sec:app}. In some cases, a third phase can appear between the
rough and flat surfaces, namely a pre-rough surface with hidden
topological order\footnote{In connection with the following discussion:
the pre-rough phase will have sound modes (the orientational order
remains unchanged compared to the rough phase), however, the
particle-hole continuum will disappear because the translational order
is different.}. So, the possible phases in the order of their
appearance are:
\begin{equation}
{\rm Locked} ~ \Longleftrightarrow ~ {\rm Rough} ~
\Longleftrightarrow ~ {\rm Pre-rough} ~ \Longleftrightarrow ~ 
{\rm Flat Faceted} ~~~ , 
\end{equation}
and these will correspond to
\begin{equation}
{\rm Ferromagnetic} \Longleftrightarrow {\rm Fermi \: Liquid}
\Longleftrightarrow {\rm Non-Fermi \: Liquid} \Longleftrightarrow ~ 
{\rm Condensate} ~~~ . 
\end{equation}
The corresponding transitions are: {\it i}) first or second-order from
the frozen solid to rough phase.  {\it ii}) Kosterlitz-Thouless or
Pokrovsky-Talapov melting between the rough  and pre-rough surfaces.
{\it iii}) first or second-order between the flat faceted and pre-rough
phases. Such a scenario predicts for two-dimensional models (e.g., the
Hubbard model) a non-Fermi liquid behavior close to half filling, that
is between the low-density Fermi liquid behaviour and the half-filled
band case. 

In regards to the rough phase, a true analysis of its properties
requires a discussion of its dynamics. In the Fermi liquid theory there
are the sound modes and also solutions with imaginary frequency that
represents the particle-hole continuum.  Immediately, it can be seen
that, the zero sound corresponds to the bosonic excitations of the
one-dimensional system. In a one-dimensional system, we would not
expect a continuum to appear because of reduced phase space. However,
treating the Fermi surface as an effective equilibrium surface we can
attach separate orientational and translational degrees of freedom, as
demonstrated in the previous Section. Such that the orientational
bosonic fluctuations are the sound modes, while the translational
fluctuations can generate a particle-hole continuum.  Being a rough
phase, these fluctuations are Gaussian, in a discrete
system\footnote{As already presented in Section\ \ref{sec:FS}, at large
length scales the discreteness of the step heights is irrelevant, due
to which the surface always can be described by a Gaussian model,
having the properties of a transverse vibrating elastic network.} the
Lagrangian is: $(g / 4 \pi) \int \vert \nabla \phi \vert^2$, with $\phi
= \pi \ell / 2$ and $\ell = \pm 1$.  We can identify this with the
six-vertex model (see, Appendix) in the following way:  in the
eight-vertex model (Baxter 1990) the magnetic charge attached to the
vertices seven and eight is $m = \pm 1$, due to which the scaling
dimension of a vortex operator is $g m^2 / 2 = g / 2$ and, accordingly,
the scaling dimension of its corresponding fugacity is $2 - g m^2 / 2 =
2 - g / 2$. If the Boltzmann weights of vertices seven and eight are
vanishingly small, the free energy is singular with an exponent inverse
proportional to the fugacity which identified with the singular
behaviour obtained by Baxter (Baxter 1990) gives :  $\cos (g \pi / 4) =
- \cos \mu$, where $\cos \mu$ is the anisotropy parameter of the model
(see, Appendix). Thus, the translation fluctuations of the Fermi
surface will have a ground-state corresponding to that of the planar
six-vertex model (Baxter 1990):
\begin{equation}
\omega_{0} ~ = ~ {\frac{\cos \mu}{4}} ~ - ~ {\frac{\sin \mu}{2}} \:
\int^{+\infty}_{- \infty}
dx \: {\frac{ \sinh (\pi - \mu) x }{ \sinh \pi x \: \cosh \mu x }} ~~~ ,
\end{equation}
with low lying states forming a particle-hole continuum bounded by
\begin{equation}
{\frac{\pi}{2}} \: {\frac{\sin \mu}{\mu}} \: | \sin k | ~ \le ~ \omega ~ \le ~
\pi \: {\frac{\sin \mu}{\mu}} \Big| \sin {\frac{k}{2}} \Big| ~~~ .
\end{equation}
Thus, not only bosons corresponding to zero sound will appear on the Fermi
surface but also a particle-hole continuum.

\section{Acknowledgments}
\label{sec:ack}

I would like to thank P. W. Anderson,  K. S. Bedell, A. R. Bishop, 
D. Coffey, J. Engelbrecht, F. D. M. Haldane, D. Scalapino and S. Trugman 
for many useful discussions.

This work was supported by the Australian Research Council.

\newpage
\section{Appendix: The Pokrovsky-Talapov Transition}
\label{sec:app}

In the past twenty years, incommensurate periodicities have been
observed in many physical systems and have become the subject of
considerable theoretical interest. Two incommensurate structures in
contact can overcome their rigidity and form a common periodicity. This
transition is known as the commensurate-incommensurate, or lock-in
transition.

Since its discovery in the context with dislocation theory (Frank and
Merwe, van der 1949), the commensurate-incommensurate transition had 
a major role (for a review, see Bak 1982; or Nijs, den 1988) in
understanding the physics of charge-density-waves, incommensurate
lattice problems, the general problem of incommensurate periodicities,
two-dimensional (XY) models, theory of phase transitions in magnetic
systems and 1D quantum systems, and the correlation driven metal
insulator (Mott) transition (Gulacsi and Bedell 1994b).

In most of the systems and models mentioned above, the problem of
characterizing the commensurate-incommensurate transition has been
reduced to the solution of the 1D sine-Gordon model. The level of
incommensurability is measured by a soliton density, $n_{\rm sol.}
(\mu)$, where $\mu$ is the chemical potential.  The
commensurate-incommensurate transition happens at $\mu = E_{\rm sol.}$,
where $E_{\rm sol.}$ is the single soliton energy. For $\mu < E_{\rm
sol.}$ and $n_{\rm sol.} = 0$ the system is locked in a commensurate
phase, while for $\mu > E_{\rm sol.}$, i.e., $n_{\rm sol.} \ne 0$ the
structure is incommensurate.

In the classical limit the commensurate-incommensurate transition
occurs with the emergence of a soliton lattice in the soliton gap
(Frank and Merwe, van der 1949; and Horowitz, 1982).  At low soliton
density, the solitons repel each other, with an exponential type
interaction (Horowitz 1982), and the transition takes place with a
logarithmic behavior $n_{\rm sol.} \propto \ln^{-1} (\mu - E_{\rm
sol.})$, i.e., corresponding to a critical exponent ${\bar \beta} = 0$.
In the full quantum problem, it is known (Haldane 1982), that the
soliton lattice will melt, giving rise to a Luttinger liquid which
emerges at the top or bottom of the soliton gap. In this case the
solitons will repel each other much stronger than in the classical
limit, with an interaction proportional to $n^2_{\rm sol.}$, in the low
soliton density limit. The incommensurability close to the transition
behaves as $n_{\rm sol.} \propto (\mu - \Delta)^{1/2}$, where $\Delta$
is the soliton gap, corresponding to a critical exponent ${\bar \beta}
= 1/2$. This is the Pokrovsky-Talapov exponent (Pokrovsky and Talapov
1978).  In some applications this transition is characterized with the
variation of the thermodynamic potential, $\Omega = E(n_{\rm sol.}) -
\mu n_{\rm sol.}$, for which we obtain $\Omega \propto n^3_{\rm sol.} =
(\mu - \Delta)^{3/2}$. From the experimental point of view, the
relevant quantity to observe this transition is the charge
susceptibility, $\chi_{\rm sol.} = (\partial \mu / \partial n_{\rm
sol.})^{-1}$, which is divergent at the transition $\chi_{\rm sol.}
\propto 1/n_{\rm sol.}$.  This divergent behavior implies a divergence
of the effective mass of the charge carriers, and accordingly the
divergence of the specific heat coefficient.  Thus, the basic
characteristics of the Pokrovsky-Talapov transition are:
\begin{equation}
n_{\rm sol.} (\mu) \propto (\mu - \Delta)^{1/2} ~ , ~~
\Omega (\mu) \propto (\mu - \Delta)^{3/2} ~ , ~~
\chi_{\rm sol.} (\mu) \propto (\mu - \Delta)^{-1/2} ~~~ .
\label{app-egy}
\end{equation}
Or, as a function of the soliton density (incommensurability) are:
\begin{equation}
\mu (n_{\rm sol.}) \propto \Delta + n^2_{\rm sol.} ~ , ~~
E (n_{\rm sol.}) \propto \Delta n_{\rm sol.} + n^3_{\rm sol.} / 2 ~ , ~~
\chi (n_{\rm sol.}) \propto 1/ n_{\rm sol.} ~~~ .
\label{app-ketto}
\end{equation}

The Pokrovsky-Talapov transition, however, is typical for all
integrable models. Its importance in connection to the 2D XY, Gaussian,
Coulomb-Gas models but, most important to the equivalent (Chui and
Weeks 1976; Beijeren, van, 1977; and Jayaprakash and Saam 1984)
solid-on-solid surface models led a to thorough analysis of its
properties.  Recognizing (Nijs, den 1981; and B. Horowitz, Bohr,
Kosterlitz and Schulz 1983) the equivalence between the Gaussian models
and the direct field six-vertex model (defined below) led to the
remarkable result that the six-vertex model in a direct field is an
ideal model to describe the commensurate-incommensurate transition. The
direct field in the six-vertex model correspond to the chemical
potential in Eqs.\ (\ref{app-egy}) and (\ref{app-ketto}), while the
polarization of the six-vertex model is equivalent to the soliton
density, $n_{\rm sol.}$.

The five- and six-vertex models are the simplest, exactly solvable
model describing surface transitions. In general, the vertex models
define a statistical mechanics on a (for simplicity) square lattice as
follows. On each link there is a degree of freedom taking two values
which is represented (traditionally) by an arrow, i.e. $\rightarrow$
and $\leftarrow$.  To each vector therefore there corresponds 16
possible configurations. To each of these we associate a Boltzmann
weight $\omega_i = \exp (- \varepsilon_i / k_B T)$. This general model
is referred to as the sixteen-vertex model. The exact solution of this
general ferroelectric model is not known. If we choose the Boltzmann
weights, so that six vertices with two entering and two exiting arrows
(or eight vertices including also the cases where four or zero arrows
exit) have finite weight, then we define the six- or eight-vertex
models. The six-vertex model is exactly solvable via the algebraic
Bethe Ansatz, for details see Lieb and Wu (Lieb and Wu 1972) and Baxter
(Baxter 1990).  The exact solution has at most four independent
variables (this is because the zero energy can be chosen arbitrary and
the ice rule implies $\omega_5 = \omega_6$)\footnote{We use the
traditional ordering of the vertices (Baxter 1990).}.  In this
situation the model is called the asymmetric six-vertex model, in
contrast to the symmetric six-vertex model, where the number of
independent variables are two. The symmetric model in the presence of
$h$ and $v$ external fields is equivalent to the asymmetric one. The
energies of the six possible arrow configurations of the symmetric
model are $\varepsilon_1 = \varepsilon_2$, $\varepsilon_3 =
\varepsilon_4$ and $\varepsilon_5 = \varepsilon_6$.  Similarly, the
field-free five-vertex model is defined by three independent Boltzmann
weights, e.g., $\varepsilon_1$, $\varepsilon_2 = \infty$,
$\varepsilon_3 = \varepsilon_4$ and $\varepsilon_5 = \varepsilon_6$ and
its exact solution can be determined via the algebraic Bethe Ansatz
(Gulacsi, Beijeren, van and Levi 1993).

{\bf 1.)} ~ For the six-vertex model, the exact free energy obtained
with the algebraic Bethe Ansatz (Lieb and Wu 1972; and Baxter 1990)
exhibits three different analytic forms for the three different phases,
i.e., two frozen-in ferroelectric phases (with the polarization, $p =
\pm 1$), an antiferroelectric phase ($p = 0$) and a disordered phase
(paraelectric, $p = 0$). The transitions to the ferroelectric phases is
of first-order, while the transition from the paraelectric phase to the
antiferroelectric one is of Kosterlitz-Thouless type.

Introducing a vertical field, $v$ in the symmetric model is still
easily solvable with the algebraic Bethe Ansatz. All the above
mentioned transitions are {\it second-order} now, as a function of
temperature and Pokrovsky-Talapov as function of the external field.
The largest eigenvalue of the transfer matrix is determined from the
Bethe Ansatz, $z(p)$ in notation of (Lieb and Wu). The exact free
energy is obtained as $ - F / k_B T$ = ${\rm max} \:  z(p)_{\: -1 \le p
\le 1}$.  All the three transitions are similar in nature, accordingly
we describe only the antiferroelectric one, $p \rightarrow 0$.  The
free energy in the ordered phase (smooth faceted surface) is determined
by (Eq.\ (304) from Lieb and Wu 1972): 
\begin{equation}
z(p) ~ \propto ~ z^{\prime} (0) \: p ~ + ~
{\frac{1}{6}} \: z^{\prime \prime \prime} (0) \: p^3  ~ + ~ \ldots ~~~ ,
\label{app-harom}
\end{equation}
with $z^{\prime \prime} = 0$, i.e., there is no second-order
contribution.  Eq.\ (\ref{app-harom}) is identical to
Eq.\ (\ref{app-ketto}). Close to the transition $p \propto (v - k_B T
\vert z^{\prime} (0) \vert)^{1/2}$ (Eq.\ (340) from Lieb and Wu 1972),
$\Omega \propto (v - k_B T \vert z^{\prime} (0) \vert)^{3/2}$
(Eq.\ (342) from Lieb and Wu 1972) and the magnetic susceptibility
$\chi \propto (v - k_B T \vert z^{\prime} (0) \vert)^{-1/2}$ which is
equivalent to Eq.\ (\ref{app-egy}). The same result is obtained for the
full asymmetric six-vertex model, i.e., with both $h$ and $v$ being
present, however, due to the complexity of the resulting equations
numerical analysis is required.

In the disordered phase (rough surface), however, the free energy has the
form (Eq.\ (316) from Lieb and Wu 1972):
\begin{equation}
z(p) ~ \propto ~  {\frac{\mu - \pi}{4}} \: \cos ({\frac{\pi \Psi_0}{2
\mu}}) \: p^2 ~
+ ~ \ldots ~~~ ,
\label{app-negy}
\end{equation}
where $\cos \mu$ is the anisotropy parameter of the model and $\Psi_0 =
(1 + \omega_1 e^{i \mu}) / (e^{i \mu} + \omega_1)$.  In connection with
Section\ \ref{sec:FS} it is important to emphasize the fact that no
linear term appears in Eq.\ (\ref{app-negy}).  In this part of the
phase diagram (i.e., on the rough surface) Taylor expansions cannot be
performed, as compared to Eq.\ (\ref{app-harom}), result of the Eq.\
(\ref{app-negy}) or (\ref{app-hat}) type are obtained directly from an
integral equation.

{\bf 2.)} ~ In the five-vertex model, in contrast to the symmetric
six-vertex model, the arrow-reversal symmetry is broken. Again there
are three different analytic forms of the exact free energy (Gulacsi,
Beijeren, van and Levi 1993) corresponding to two frozen-in
ferroelectric phases ($p = \pm 1$), one antiferroelectric phase ($p =
0$) (similar to the symmetric six-vertex model) and a ferrielectric
($-1 \le p \le 1$) phase.  All the transitions are of second-order.
From all these phases, the ferrielectric and ferroelectric  phases are
connected to surfaces (see later) due to which I briefly present their
critical behaviour. In this case, e.g., $p \rightarrow 1$ and the free
energy in the ordered phase (smooth facets) is (Eq.\ (A7) and (A8) from
Gulacsi, Levi and Tosatti 1994):
\begin{eqnarray}
f (1 - p) ~ \propto ~ f(0) ~ & + & ~ k_B T \: \Big[
{\frac{\omega_1}{\omega_1 - \omega_3}}
\ln \Big( {\frac{\omega_1}{\omega_3}}\Big) + 2 \ln
\Big({\frac{\omega_3}{\omega_5}}\Big)
\Big]  \: {\frac{1 - p}{2}}  \cr
& - & ~ k_B T \: {\frac{\pi^2}{6}} \: {\frac{\omega_1 \omega_3 (\omega_1 +
\omega_3)}
{(\omega_1 - \omega_3)^3}} \: \ln \Big( {\frac{\omega_1}{\omega_3}}\Big)
\: \Big( {\frac{1 - p}{2}} \Big)^{3}~ + ~  \ldots ~~~ .
\label{app-ot}
\end{eqnarray}
As in the six-vertex model, see Eq.\ (\ref{app-harom}) there is no
second-order term.  Close  to the transition $1 - p$, $\Omega$ and the
susceptibility behaves similar to the previously presented six-vertex
case or Eq.\ (\ref{app-egy}). The free energy on the ferrielectric side
(rough surface) has a particular simple form in the non-interacting,
zero external field limit:
\begin{equation}
f (1 - p) ~ \propto ~ f(0) ~ + ~ \pi^2 k_B T \: \Big( {\frac{1 -
p}{2}}\Big)^2 ~ + ~ \ldots ~~~ ,
\label{app-hat}
\end{equation}
while for the full formula see Eq.\ (21) from Gulacsi {\sl et al.}
(Gulacsi, Levi and Tosatti 1994). Because of the simplicity of
Eq.\ (\ref{app-hat}) the five-vertex model is most adequate to model
the Pokrovsky-Talapov transition. 

As a third example, I present briefly a variant of the
Pokrovsky-Talapov transition appearing on surfaces, similar to
Section\ \ref{sec:facet}, known also as  Gruber-Mullins (1967)
transition.  Surfaces of simple cubic crystals are described by the
five-vertex model (Gulacsi, Levi and Tosatti 1994), while surfaces of
body centered  crystals with the six-vertex model (Beijeren, van, 1977;
and Jayaprakash and Saam 1984) As presented in Section\ \ref{sec:surf},
knowing the free energy of the surface we can determine the Cartesian
coordinates, $z(x,y)$, of the corresponding surface from the Wulff
construction.  The free energy of the surface we obtain directly from
the exact solution of the five- or six-vertex model, depending on the
crystal structure.  For the simple cubic surfaces, $p \rightarrow 1$
and $1 - p \propto (x - x_0)^{1/2}$, while for surface of body centered
crystals $p \rightarrow 0$ and $p \propto (x - x_0)^{1/2}$. In both
cases, $z - z_0 \propto (x - x_0)^{3/2}$ and the average spacing
between steps $\propto (x - x_0)^{-1/2}$, equations identical to
Eq.\ (\ref{app-egy}).

As mentioned above, a general characteristic of the (Luttinger) liquid
phase preceding the Pokrovsky-Talapov transition is that its
quasiparticles strongly repel each other. The basic equation obtained
for the ground-state energy in Eq.\ (\ref{app-ketto}) can be
interpreted as follows: the linear term in $n_{\rm sol.}$ sums up the
energy contributions of the non-interacting quasiparticles (solitons,
holons, surface steps, etc., depending on the model under
consideration).  The missing quadratic term in $n_{\rm sol.}$ shows
that there is no effective {\it attraction} between the quasiparticles
(see below) and the $n^3_{\rm sol.}$ term suggests of the presence of
strong repulsion.  This repulsion is discussed in detail in
Section\ \ref{sec:facet}.

As all the previous proof were given for an effective nearest-neighbour
interacting systems, it is very important to realize that the
Pokrovsky-Talapov transition is very stable with respect to different
types of interactions. Short range interactions, with range much less
than the average distance between the quasiparticles contribute to the
ground-state energy by an order of $n^4_{\rm sol.}$ thus, leaving the
Pokrovsky-Talapov transition unaltered as it was proven in
Section\ \ref{sec:facet}, where also the cases of long-range
interactions was analyzed.

By the above examples we want to emphasize that the integrable models
have common ground-state properties\footnote{It should be emphasized
that the above mentioned mappings between different models establishes
an equivalence between the eigenvalues of the different models. That
is, the ground-state energies and thus, the thermodynamic properties of
the different models can be made equivalent. However, non of these
mapping schemes preserves the correlation functions.} thus, a
Pokrovsky-Talapov type transition will exist in all related models.

\newpage
\section{References}

Andreev, A. F. and Pashkin, A. Ya, 1978, Sov. Phys. JETP, {\bf 48}, 763.

Ashcroft, N. W. and Mermin, N. D., 1976, {\it Solid state Physics},
(Holt, Rinehart and Winston, New York) p. 350.

Bak, P., 1982, Rep. Prog. Phys., {\bf 45}, 587.

Bauer, E., 1987, in {\it Structure and Dynamics of Surfaces}, eds.
W. Schommers and P. von Blanckenhagen, Vol. 2, Topics in Current Physics, 
Vol. 43, (Springer-Verlag, Berlin) p. 115. 

Balibar S., Guthmann C. and Rolley E., 1993, J. Phys. I(Paris), {\bf 3}, 1475.

Baxter R. J., 1990, {\it Exactly Solvable Models in Statistical Mechanics},
(Academic  Press, New York). 

Beijeren H. van, 1977, Phys. Rev. Lett., {\bf 38}, 993.

Beijeren, H. van and Nolden I., 1987, in {\it Structure and Dynamics of
Surfaces}, eds. W. Schommers W. and P. von Blanckenhagen, Vol. 2, Topics 
in Current Physics, Vol. 43, (Springer-Verlag, Berlin) p. 259. 

Burton, W. K. and Cabrera, N., 1949, Disc. Faraday Soc. {\bf 5}, 40

Burton, W. K., Cabrera, N. and Frank, F. C., 951, Phil. Trans. Roy. Soc.,
A{\bf 243}, 299.

Chui S. T. and Weeks J. D., 1976, Phys. Rev., B{\bf 14}, 4978.

Fisher, D. S. and Weeks, J. D., 1983, Phys. Rev. Lett., {\bf 50}, 1077.

Fradkin, E., 1983, Phys. Rev., B{\bf 28}, 5338.

Frank, F. C.  and Merwe, J. H. van der, 1949,  Proc. Royal Soc., A{\bf
198}, 205.

Gruber E. E. and Mullins W. W., 1967, J. Phys. Chem. Solids, {\bf 28}, 875.

Gulacsi M, Beijeren H. van and Levi, A. C., 1993, Phys. Rev., E{\bf 47}, 2473.

Gulacsi M., Levi A. C. and Tosatti, E., 1994a, Phys. Rev., E{\bf 49}, 3843.

Gulacsi M. and Bedell K. S., 1994b, Phys. Rev. Lett., {\bf 72}, 2765.

Gulacsi, M., 1996, in preparation.

Haldane, F. D. M., 1982, J. Phys., A{\bf 15}, 507.

Haldane, F. D. M., 1994, in {\it Perspectives in Many-Particle Physics},
eds. R. A. Broglia, J. R. Schrieffer and P. F. Bortignon, (North-Holland,
Amsterdam) p. 5. 

Herring, C., 1951, Phys. Rev., {\bf 82}, 368.

Honner, G, 1995, private communication.

Horowitz, B., 1982, J. Phys., C{\bf15}, 161.

Horowitz B., Bohr T., Kosterlitz J. M. and Schulz H. J., 1983, Phys. Rev.,
B{\bf 28}, 6596.

Jayaprakash C. and Saam W. F., 1984, Phys. Rev., B{\bf 30}, 3917.

Jayaprakash C., Rottman C. and Saam W. F., 1984, Phys. Rev., B{\bf 30}, 6549.

Kosterlitz, J. M. and Thouless D. J., 1973, J. Phys., C{\bf 6}, 1181.

Lieb E. H. and Wu F. Y., 1972, in {\it Phase Transition and critical
Phenomena}, eds. C. Domb and N. S. Green, Vol. 1, (Academic Press,
New York) p. 332. 

Landau, L. D., 1957, Sov. Phys. JETP, {\bf 3}, 920 and 1957, {\sl ibid.},
{\bf 5}, 101.

Landau, L. D. and Lifshitz, E. M., 1980, {\it Statistical Physics}, Vol. 1,
(Pergamon Press, Oxford) p. 517.

Nijs M. P. M. den, 1981, Phys. Rev., B{\bf 23}, 6111.

Nijs M. P. M. den, 1988, in {\it Phase Transition and critical Phenomena},
eds. C. Domb and N. S. Green, Vol. 12, (Academic Press, New York) p. 220. 

Nozi\`{e}res, P., 1991, Lect. Notes of the 1989 Beg-Rohu Summer School, 
in {\it Solids Far From Equilibrium}, ed. C. Godreche, (Cambridge University Press, cambridge) p. 1. 

Pines, D. and Nozi\`{e}res, P., 1988, {\it The Theory of Quantum Liquids},
(Addison-Wesley, New York) p. 15.

Pokrovsky V. L. and Talapov A. L., 1978, Sov. Phys. JETP,  {\bf 48}, 579;
1979, Phys.Rev. Lett.,
{\bf 42}, 65; and 1980, Sov. Phys. JETP, {\bf 51}, 134.

Pomeranchuk, Y. A., 1958, Sov. Phys. JETP, {\bf 8}, 361. 

Rottman, C. and Wortis, M. 1984, Phys. Rev. B{\bf 29}, 328; and Phys. Rep. 
{\bf 103}, 59. 

Wulff, G., 1901, Z. Kritallogr. Mineral., {\bf 34}, 449.

\section{Additional References for Surface Physics}

Domb, C. and Lebowitz, J. L., eds., 1986, {\it Phase Transition and
Critical Phenomena}, Vol. 9, 10, 12, Academic.

Blakely, J. M., ed., 1975, {\it Surface Physics of Materials}, Vol. 1, Academic.

Bruce, A. D. and Cowley, R. A., 1981, {\it Structural Phase Transitions},
Taylor and Francis.

Cohen, E. G. D., ed., 1985, {\it Fundamental Problems in Statistical
Mechanics}, Vol. 6, Elsevier.

Dash, J. G. and Ruvalds, J., eds., 1980, {\it Phase Transition in Surface
Films}, Plenum.

Gomer, R. and Smith, G. S., eds., 1953, {\it Structure and Properties of
Solid Surfaces}, Univ.
of Chicago Press.

Hahne, F. W., ed., 1983, {\it Critical Phenomena}, Lect. Notes in Physics,
Vol. 186, Springer.

Kingston, W. E., ed., 1951, {\it Physics of Powder Metallurgy}, McGraw-Hill.

Patashinskii, A. Z. and Pokrovsky, V. L., 1979, {\it Fluctuation Theory of
Phase Transitions},
Pergamon.

Pekalski, A. and J. Sznajd, eds., 1984, {\it Static Critical Phenomena in
Inhomogeneous
Systems}, Lecture Notes in Physics, Vol. 206, Springer.

Riste, T., ed., 1980, {\it Ordering in Strongly Fluctuating Condensed
Matter Systems}, Plenum.

Schommers W. and Blanckenhagen, P. von, eds.,1987, {\it Structure and
Dynamics of Surfaces},
Vol. 2, Topics in Current Physics, Vol. 43, Springer.

Sinha, K., ed.,1980, {\it Ordering in Two Dimension}, North Holland.

Stanley, H. E., 1971, {\it Introduction to Phase Transition and Critical
Phenomena},  Claredon.

\newpage

\section{Figure Caption}

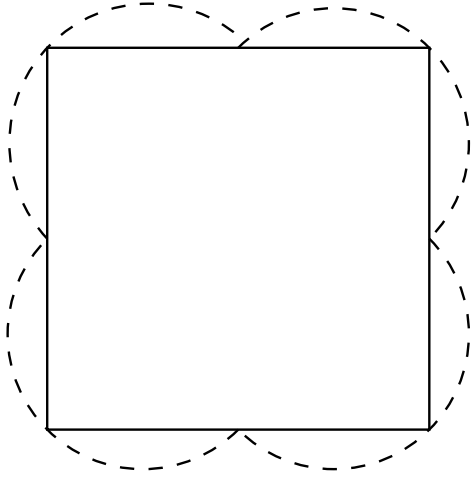
Fig. 1. ~ The surface of a square (solid line) and its corresponding Wulff
plot (dashes line).

\end{document}